\documentclass[pra,aps,amsmath,amssymb,showkeys,showpacs]{revtex4}
\usepackage{graphicx}
\newcommand{\hh}{\mathcal{H}}
\newcommand{\lnp}{{\mathcal{L}}}
\newcommand{\lsa}{{\mathcal{L}}_{s.a.}}
\newcommand{\lsp}{{\mathcal{L}}_{+}}
\newcommand{\cla}{\mathcal{A}}
\newcommand{\clb}{\mathcal{B}}

\newcommand{\zset}{\mathbb{C}}
\newcommand{\mset}{\mathbb{M}}
\newcommand{\rset}{\mathbb{R}}
\newcommand{\pen}{\openone}
\newcommand{\bro}{\boldsymbol{\rho}}

\newcommand{\bau}{\boldsymbol{\tau}}
\newcommand{\zmx}{\mathbf{0}}
\newcommand{\am}{{\mathsf{A}}}
\newcommand{\bn}{{\mathsf{B}}}
\newcommand{\cn}{{\mathsf{C}}}
\newcommand{\enn}{{\mathsf{E}}}
\newcommand{\gm}{{\mathsf{G}}}
\newcommand{\psf}{{\mathsf{\Pi}}}
\newcommand{\lsf}{{\mathsf{\Lambda}}}
\newcommand{\mn}{{\mathsf{M}}}
\newcommand{\ppm}{{\mathsf{P}}}
\newcommand{\qpm}{{\mathsf{Q}}}

\newcommand{\um}{{\mathsf{U}}}
\newcommand{\vm}{{\mathsf{V}}}
\newcommand{\wm}{{\mathsf{W}}}
\newcommand{\ax}{{\mathsf{X}}}
\newcommand{\ay}{{\mathsf{Y}}}
\newcommand{\az}{{\mathsf{Z}}}
\newcommand{\gmsf}{{\mathsf{\Gamma}}}
\newcommand{\simf}{{\mathsf{\Sigma}}}
\newcommand{\Tr}{{\mathrm{Tr}}}
\newcommand{\id}{{\mathrm{id}}}
\newcommand{\ppc}{{\mathcal{P}}}
\newcommand{\qpc}{{\mathcal{Q}}}
\newcommand{\wtc}{\widetilde{c}}
\newcommand{\wto}{\widetilde{\omega}}

\begin{document}
\clearpage
\preprint{}

\title{Majorization entropic uncertainty relations for
                   quantum operations}

\author{Alexey E. Rastegin$^{1}$ and Karol \.{Z}yczkowski$^{2,3}$}
\affiliation{
\mbox{$^1$Department of Theoretical Physics, Irkutsk State University, Gagarin Bv. 20,
Irkutsk 664003, Russia}\\
\mbox{$^2$Institute of Physics, Jagiellonian University,
ul.\ {\L}ojasiewicza 11, 30-348 Krak\'ow, Poland}\\
\mbox{$^3$Center for Theoretical Physics, Polish Academy of Sciences, al.\ Lotnik\'ow 32/46, 02-668 Warszawa, Poland}
}

\begin{abstract}
Majorization uncertainty relations are derived for arbitrary
quantum operations acting on a finite-dimensional space. The basic
idea is to consider submatrices of block matrices comprised of the
corresponding Kraus operators. This is an extension of the
previous formulation, which deals with submatrices of a unitary
matrix relating orthogonal bases in which measurements are
performed. Two classes of majorization relations are considered:
one related to tensor product of probability vectors and another
one related to their direct sum. We explicitly discuss an example
of a pair of one-qubit operations, each of them represented by two
Kraus operators. In the particular case of quantum maps describing
orthogonal measurements the presented formulation reduces to
earlier results derived for measurements in orthogonal bases. The
presented approach allows us also to bound the entropy
characterizing results of a single generalized measurement.
\end{abstract}
\pacs{03.65.Ta, 03.67.-a, 03.67.Ud}
\keywords{majorization, quantum operation, R\'{e}nyi entropy, Tsallis entropy}

\maketitle

\pagenumbering{arabic}
\setcounter{page}{1}

\section{Introduction}\label{sec1}

The Heisenberg uncertainty relation \cite{wh27} is one of the
fundamental limitations in description of the quantum world. It is
commonly emphasized that one is unable to assign simultaneously
precise  values to non-commuting observables. Despite a vast
literature concerning the uncertainty relations, there is an on
going debate over their scope and validity \cite{lahti,hall99}.
The traditional formulation of Robertson \cite{robert} gives a
lower bound on the product of the standard deviations of two
observables. This state-dependent bound is proportional to the
modulus of the expectation value of the commutator in the measured
state. Dependence on the pre-measurement state results in the fact
that Robertson's bound may vanish, even if the commutator itself
is non-zero. For instance, a trivial bound is obtained for any
eigenstate of any of the two observables \cite{maass}. Relations
with a lower bound on product of the standard deviations are
useful for a small class of quantum observables \cite{zimba00}.

There are several other approaches to measure the amount of
uncertainty in quantum measurements. Entropic uncertainty
relations are currently the subject of a considerable scientific
interest \cite{ww10,brud11,cbtw15}. For the case of canonically
conjugate variables, this approach was initiated by Hirschman
\cite{hirs} and later developed in \cite{beck,birula1}. Advantages
of entropic uncertainty relations for finite-dimensional systems
were emphasized in \cite{deutsch,kraus87}. The famous result of
Maassen and Uffink \cite{maass} is based on the Riesz theorem and
inspires many later formulations. Entropic uncertainty
relations were derived for several special measurements such as
mutually unbiased bases \cite{ivan92,sanchez93,molm09a}. This
question was further examined with use of both the generalized
entropies of R\'{e}nyi and Tsallis \cite{rastmub}. Another
viewpoint on the Maassen--Uffink bound is connected with the role
of a quantum memory \cite{BCCRR10}. Improved entropic uncertainty
relations in the presence of quantum correlations were obtained in
\cite{coles14}.

As was emphasized in \cite{oppwn10}, entropic bounds cannot
distinguish the uncertainty inherent in obtaining a particular
combination of the outcomes. Studies of the fine-grained
uncertainty relations were initiated in \cite{oppwn10} and further
developed in \cite{renf13,rastqip15,Rud15}. Majorization relations
offer an alternative way to express the uncertainty principle in
terms of probabilities {\sl per se} \cite{prtv11}. At the same
time, majorization relations directly lead to desired inequalities
in terms of the Shannon entropy or other generalized entropies.
First majorization entropic uncertainty relations were based on
tensor product of probability vectors \cite{prz13,fgg13}. Even
stronger bounds obtained in \cite{rpz14} are based on majorization
relations applied to direct-sum of probability vectors.

In this work we extend the techniques applied in
\cite{prz13,rpz14} for orthogonal projective measurements for any
trace-preserving quantum operations in finite dimensions. This
issue was already discussed in \cite{fgg13}, but our present
results go much further and lead to explicit bounds easy to apply
in the general case. The case of POVM measurements is included
into discussion, as quantum measurements can be described in terms
of Kraus operators. In our approach we use the majorization
techniques for vectors depending on norms of submatrices. In the
case of orthogonal measurements, the authors of \cite{prz13,rpz14}
used submatrices of a unitary matrix, relating both bases. In this
work, concerning the general case of arbitrary quantum operations,
we consider submatrices of block matrices formed by concatenation
of Kraus operators, which represent the quantum operation.
Majorization uncertainty relations obtained in this work are
compared with generalizations of the Maassen--Uffink bounds for
quantum operations \cite{KP902,rast104}.

The paper is organized as follows. In Sect. \ref{sec2}, we review
required notions and preliminary facts of matrix analysis.
Previous results on majorization uncertainty relations are briefly
discussed in Sect. \ref{sec3}. The majorization uncertainty
relations for quantum operations are obtained in Sect. \ref{sec4}.
In Sect. \ref{sec5} we consider an example of two quantum
operations acting on a single qubit and represented by two Kraus
operators each. Majorization bounds derived here are compared with
uncertainty relations of the Maassen--Uffink type. In Sect.
\ref{sec6}, we show that in the case of two orthogonal
measurements the proposed formulation reduces to the former one.
The paper is concluded in Sect. \ref{sec7} with a brief summary of
results obtained. In \ref{apn1}, we apply of the
majorization approach to the case of a single quantum operation.

\section{Preliminaries}\label{sec2}

In this section, preliminary facts are briefly outlined. For any
two integers $m,n\geq1$ the symbol  $\mset_{m\times{n}}(\zset)$
denotes the space of all $m\times{n}$ complex matrices
\cite{hiai2014}. If $m=n$, we will usually write
$\mset_{n}(\zset)$. Each matrix $\ax\in\mset_{m\times{n}}(\zset)$
can be represented in terms of the singular value decomposition
\cite{hj1990},
\begin{equation}
\ax=\um\simf\vm^{\dagger}
\ , \label{svdx}
\end{equation}
where $\um\in\mset_{m}(\zset)$ and $\vm\in\mset_{n}(\zset)$ are
unitary. The matrix
$\simf=[[\sigma_{ij}]]\in\mset_{m\times{n}}(\zset)$ has
$\sigma_{ij}=0$ for all $i\neq{j}$. If the given matrix $\ax$ has
rank $k$, then we can write $\simf$ so that
\begin{equation}
\sigma_{11}\geq\cdots\geq\sigma_{kk}>\sigma_{k+1,k+1}=\cdots=\sigma_{rr}=0
\ , \label{sngv}
\end{equation}
where $r=\min\{m,n\}$. The diagonal entries
$\{\sigma_{jj}\}=\{\sigma_{j}(\ax)\}$ of $\simf$ are known as the
singular values of $\ax$. The non-zero $\sigma_{jj}$'s are the
squared roots of positive eigenvalues of both $\ax\ax^{\dagger}$
and $\ax^{\dagger}\ax$. A coordinate-independent formulation of
the above facts can be found in \cite{watrous1}.

We will use the notion of the spectral norm of a matrix, which can
be written in terms of the singular values,
\begin{equation}
\|\ax\|_{\infty}=\max\bigl\{\sigma_{j}(\ax):{\>}1\leq{j}\leq{r}\bigr\}
\ .
\label{spnm}
\end{equation}
It is sometimes referred to as the operator norm. Other norms are
widely used in quantum information theory \cite{watrous1} for
quantifying various properties of a matrix. Many of them are also
defined in terms of singular values. For instance, the Schatten
norms form an important family of unitarily invariant norms. A
norm $||\centerdot||$ is said to be unitarily invariant if
\begin{equation}
||\um\ax\vm||=||\ax||
\label{uinn}
\end{equation}
for all $\ax\in\mset_{m\times{n}}(\zset)$ and for all unitary
matrices $\um\in\mset_{m}(\zset)$, $\vm\in\mset_{n}(\zset)$
\cite{hj1990}. We will also exploit the following two facts. Let
$\ax$ and $\ay$ be two rectangular matrices of the same size. Then
the products $\ax\ay^{\dagger}$ and $\ay^{\dagger}\ax$ have the
same non-zero eigenvalues. If $\ax$ is a submatrix of $\ay$, then
$\|\ax\|_{\infty}\leq\|\ay\|_{\infty}$.

The space of linear operators on finite-dimensional Hilbert space
$\hh$ will be denoted as $\lnp(\hh)$. By $\lsa(\hh)$ and
$\lsp(\hh)$ we mean the real space of Hermitian operators and the
set of positive operators, respectively. Let square matrices $\ax$
and $\ay$ represent elements of $\lnp(\hh)$. Their
Hilbert--Schmidt product is defined as \cite{watrous1}
\begin{equation}
\langle\ax{\,},\ay\rangle_{\mathrm{hs}}:=\Tr(\ax^{\dagger}\ay)
\ . \label{axay}
\end{equation}
This inner product induces the Hilbert--Schmidt norm of matrices.
If the reference basis is fixed, we can avoid distinguishing
between elements of $\lnp(\hh)$ and these matrices which represent
them. We shall prove another statement concerning the spectral
norm.

\newtheorem{lem01}{Lemma}
\begin{lem01}\label{pn01}
Let $\ax$ be a rectangular matrix, and let $\ay$ be obtained from
$\ax$ by adding rows of zeros and columns of zeros. Then these
matrices have the same spectral norm, i.e.,
$\|\ax\|_{\infty}=\|\ay\|_{\infty}$.
\end{lem01}

{\bf Proof.} Without loss of generality, we assume that zero rows
are placed below and zero columns are added right. This is merely
rearranging of vectors of the standard basis. So, we can write
$$
\ay=
\begin{pmatrix}	
\ax & \zmx \\
\zmx & \zmx
\end{pmatrix}
 , 
$$
where $\zmx$'s are zero submatrices. Due to (\ref{svdx}), we further obtain
\begin{equation}
{\mathrm{diag}}(\um,\pen)
\,\ay\,{\mathrm{diag}}(\vm^{\dagger},\pen)
=
\begin{pmatrix}
\simf & \zmx \\
\zmx & \zmx
\end{pmatrix}
 . \label{axyz}
\end{equation}
By $\pen$, we mean the identity matrices of the corresponding
size. Formula (\ref{axyz}) shows that non-zero singular values
of the matrices $\ax$ and $\ay$ are the same. $\blacksquare$

To approach majorization uncertainty relations, the authors of
\cite{prz13,rpz14} inspected norms of submatrices of a certain
unitary matrix. To the given two orthonormal bases
$\{|e_{i}\rangle\}$ and $\{|f_{j}\rangle\}$, with
$i,j=1,\ldots,d$, we assign the unitary matrix $\wm$ of size $d$
with entries $w_{ij}=\langle{e}_{i}|f_{j}\rangle$. By
$\mathcal{SUB}(\wm,k)$, we mean the set of all its submatrices of
class $k$ defined by
\begin{equation}
\mathcal{SUB}(\wm,k):=
\bigl\{
\mn\in\mset_{r\times{r}^{\prime}}(\zset):{\>}r+r^{\prime}=k+1,{\>}
\mn {\text{ is a submatrix of }} \wm
\bigr\}
\, . \label{subvk}
\end{equation}
The positive integer $k$ runs all the values, for which the
following condition for sum of integer dimensions holds,
$r+r^{\prime}=k+1$. The majorization relations of
\cite{prz13,rpz14} are expressed in terms of quantities
\begin{equation}
s_{k}:=\max\bigl\{
\|\mn\|_{\infty}:{\>}\mn\in\mathcal{SUB}(\wm,k)
\bigr\}
\, . \label{skdf}
\end{equation}
This definition can be extended to an arbitrary matrix and,
furthermore, to a block matrix. Let $\ax$ be a $M\times{N}$ matrix
of blocks $\ax_{ij}$, each of size $d$, namely
\begin{equation}
\ax=
\begin{pmatrix}
\ax_{11} & \cdots & \ax_{1N} \\
\cdots & \cdots & \cdots \\
\ax_{M1} & \cdots & \ax_{MN}
\end{pmatrix}
 . \label{bmxmn}
\end{equation}
Calculations with block matrices are similar to these with
ordinary matrices \cite{hiai2014}. We should only remember that
their entries do not commute. Block submatrices of (\ref{bmxmn})
are defined quite similarly to submatrices of ordinary matrices.
That is, one fixes a subset of rows and a subset of columns in
(\ref{bmxmn}). Keeping the blocks on the points of intersection
will result in a submatrix composed of some $d\times{d}$ blocks of
$\ax$. This definition should not lead to a confusion since the
size of blocks is prescribed initially. We will now extend
definitions (\ref{subvk}) and (\ref{skdf}). Let us define the set
\begin{equation}
\mathcal{BSUB}(\ax,k):=
\Bigl\{
\az \in\mset_{rd\times{r}^{\prime}d}(\zset):{\>}r+r^{\prime}=k+1,{\>}
\az {\text{ is a submatrix of $d\times{d}$-blocks of $\ax$}}
\Bigr\}
\, . \label{bsubvk}
\end{equation}
Again, the positive integer $k$ runs all the values with respect
to $r+r^{\prime}=k+1$. For $M\times{N}$ block matrix
(\ref{bmxmn}), we have $1\leq{k}\leq{M}+N-1$. Similarly to
(\ref{skdf}), we further introduce a sequence with positive
elements
\begin{equation}
c_{k}:=\max\bigl\{
\|\az\|_{\infty}:{\>}\az\in\mathcal{BSUB}(\ax,k)
\bigr\}
\, . \label{bskdf}
\end{equation}

The aim of the present work is to extend the techniques applied in
\cite{prz13,rpz14} for orthogonal measurements for trace-preserving completely
positive maps, also called {\sl quantum
operations} or stochastic maps. Let us recall some basic notions
of the field. We will consider a linear map
$\Phi:\,\lnp(\hh)\rightarrow\lnp(\hh^{\prime})$ that takes
elements of $\lnp(\hh)$ to elements of $\lnp(\hh^{\prime})$. This
map is called positive, when elements of $\lsp(\hh)$ are mapped to
elements of $\lsp(\hh^{\prime})$ \cite{bhatia07}. To describe
physical processes, linear maps have to be completely positive
\cite{bengtsson,nielsen}. Let $\id^{\prime\prime}$ be the identity
map on $\lnp(\hh^{\prime\prime})$, where the space
$\hh^{\prime\prime}$ is assigned to a reference system. Complete
positivity implies that the map $\Phi\otimes\id^{\prime\prime}$ is
positive for any dimension of the auxiliary space
$\hh^{\prime\prime}$. A completely positive map $\Phi$ can be
written by an operator-sum representation. For the input density
matrix $\bro\in\lsp(\hh)$, the output one is then written as
\begin{equation}
\Phi(\bro)=\sum\nolimits_{i} \am_{i}\bro\,\am_{i}^{\dagger}
\ . \label{opsm}
\end{equation}
Here, the Kraus operators $\am_{i}$ map the input space $\hh$ to
the output space $\hh^{\prime}$. They are described by matrices
with ${\mathrm{dim}}(\hh^{\prime})$ rows and ${\mathrm{dim}}(\hh)$
columns. Representations of the form (\ref{opsm}) are not unique
\cite{watrous1}. Each concrete set $\{\am_{i}\}$ resulting in
(\ref{opsm}) will be referred to as ``unraveling'' of the map
$\Phi$. This terminology is due to Carmichael \cite{carm} who
introduced this notion to represent master equations. If a
physical process is closed and the probability is conserved, the
map preserves the trace:
\begin{equation}
\Tr\bigl(\Phi(\bro)\bigr)=\Tr(\bro)
\ . \label{cotr}
\end{equation}
This relation satisfied for all $\bro$ is equivalent to
the following constraint for the set of the Kraus operators:
\begin{equation}
\sum\nolimits_{i} \am_{i}^{\dagger}\am_{i}=\pen
\ . \label{prtr}
\end{equation}
Here $\pen$ denotes the identity operator on the input space $\hh$.

A general quantum measurement can be described by a collection of
measurement operators \cite{nielsen}. These measurement operators
obey the  relation (\ref{prtr}), so we can treat them as Kraus
operators. If the pre-measurement state is described by a density
matrix $\bro\in\lsp(\hh)$, normalized as $\Tr(\bro)=1$, then the
probability of $i$-th outcome is written as
$\Tr\bigl(\am_{i}^{\dagger}\am_{i}\bro\bigr)$. When we are
interested only in the probabilities of the respective outcomes,
we may restrict an attention to positive semidefinite operators
$\enn_{i}=\am_{i}^{\dagger}\am_{i}$. The set $\{\enn_{i}\}$ gives
a non-orthogonal resolution of the identity often called {\sl
positive operator-valued measure} (POVM). Thus, the POVM formalism
to deal with quantum measurements is naturally involved into our
approach. We only note that measurement operators give a more
refined description of the measurement process.

For any quantum operation $\Phi$ one has a certain freedom in the
operator-sum representation. Suppose that the set
$\{\widetilde\am_{i}\}$ is another unraveling of the quantum
operation (\ref{opsm}). Then the Kraus operators are related as
\begin{equation}
\widetilde{\am}_{i}=\sum\nolimits_{j} \gamma_{ij}{\,}\am_{j}
\ , \label{eqvun}
\end{equation}
where the matrix $\gmsf=[[\gamma_{ij}]]$ is unitary
\cite{nielsen}. Here, we assume that the sets $\cla=\{\am_{i}\}$
and $\widetilde{\cla}=\{\widetilde{\am}_{i}\}$ have the same
cardinality by adding zero operators, if needed. A size of the
matrix $\gmsf$ corresponds to this cardinality.

Finally, we recall some notions of majorization. Let
$x=(x_{1},\ldots,x_{N})$ and $y=(y_{1},\ldots,y_{N})$ be vectors
in $\rset^{N}$. The majorization ${x}\prec{y}$ implies that
\cite{hiai2014}
\begin{equation}
\sum\nolimits_{i=1}^{k}x_{i}^{\downarrow}
\leq \sum\nolimits_{i=1}^{k}y_{i}^{\downarrow}
\qquad (1\leq{k}\leq{N})
\ , \label{mrdf}
\end{equation}
and equality is required for $k=N$. Here the arrows down imply
that the components should be put in the decreasing order. As was
discussed in \cite{prtv11,prz13,fgg13,rpz14}, the uncertainty
principle may sometimes be expressed by a majorization relation.

\section{Majorization uncertainty relations for two von Neumann measurements}\label{sec3}

In this section, we briefly discuss a general formulation of
majorization uncertainty relations for two von Neumann
measurements. Definitions of the used entropic measures are
recalled as well. To each von Neumann measurement, we can assign
some orthogonal resolution of the identity. We first suppose that
measured observables are non-degenerate, whence the orthogonal
projectors are all of rank-one. It will be convenient to write the
dimensionality explicitly. Thus, we deal with two orthonormal
bases in $d$-dimensional Hilbert space $\hh_{d}$. In any
prescribed basis, vectors $|\psi\rangle\in\hh_{d}$ are represented
as elements of $\mset_{d\times1}(\zset)$.

Let us consider two orthonormal bases denoted by
$\{|e_{i}\rangle\}$ and $\{|f_{j}\rangle\}$ with $i,j=1,\ldots,d$.
If the pre-measurement state is described by the normalized
density matrix $\bro\in\lsp(\hh_{d})$, then elements of the
generated probabilistic vectors $p$ and $q$ are expressed as
\begin{align}
p_{i}&=\langle{e}_{i}|\bro|e_{i}\rangle
\ , \label{pidf}\\
q_{j}&=\langle{f}_{j}|\bro|f_{j}\rangle
\ . \label{qjdf}
\end{align}
There are several ways to formulate uncertainty relations. In this
approach, entropic functions are used to quantify an amount of
uncertainty associated with a generated probability distribution.
We will use both the R\'{e}nyi and the Tsallis entropies.

For $0<\alpha\neq1$, the R\'{e}nyi $\alpha$-entropy is defined as
\cite{renyi61}
\begin{equation}
H_{\alpha}(p):=
\frac{1}{1-\alpha}{\>}
\ln\!\left(\sum\nolimits_{i=1}^{d} p_{i}^{\alpha}
\right)
\, . \label{renent}
\end{equation}
This entropy is a non-increasing function of the R\'{e}nyi
parameter $\alpha$ \cite{renyi61}. Quantum information measures of
the R\'{e}nyi type are examined in \cite{mdsft13,berta15}. The
Tsallis $\alpha$-entropy of degree $0<\alpha\neq1$ is defined by
\cite{tsallis}
\begin{equation}
T_{\alpha}(p):=\frac{1}{1-\alpha}{\>}
\left(\sum\nolimits_{i=1}^{d} p_{i}^{\alpha}
- 1\right)
\, . \label{tsaent}
\end{equation}
Quantum applications of Tsallis entropic functions are considered
in \cite{sudha14,rast15ap,rast16a}. In the limit $\alpha\to1$,
both entropies (\ref{renent}) and (\ref{tsaent}) lead to the
Shannon entropy
\begin{equation}
H_{1}(p)=-\sum\nolimits_{i=1}^{d} p_{i}\ln{p}_{i}
\ . \label{shent}
\end{equation}
Note that for $\alpha\to1$ the Tsallis entropy tends to the
Shannon entropy if the latter quantity is defined via natural
logarithm as in (\ref{shent}). If required, one can rescale the
above formulas in order to change the base of the logarithm. In
the following, we will use the fact that entropies (\ref{renent})
and (\ref{tsaent}) are both Schur concave. Many properties
of entropies (\ref{renent}) and (\ref{tsaent}) and their
applications are discussed in the book \cite{bengtsson}.

We now recall some details of the majorization approach in finite
dimensions. Applications of this approach to the case of position
and momentum are discussed in \cite{prtv11}. Suppose $p$ and $q$
denote two probability vectors generated by two quantum
measurements performed on two copies of the same quantum state.
The key idea of this approach is to majorize some binary
combination of $p$ and $q$ by a third vector with bounding
elements. Majorization relations of the tensor-product type, first
considered in \cite{fgg13,prz13}, are obtained by finding a
probability vector $\omega^{\prime}$ such that the following
majorization relation holds,
\begin{equation}
p\otimes{q}\prec\omega^{\prime}
\ . \label{pq0w}
\end{equation}
In the present paper, we also deal with majorization relations
of the direct-sum type originally introduced in \cite{rpz14}.
This approach, usually producing stronger bounds,
is based on the relation
\begin{equation}
p\oplus{q}\prec\{1\}\oplus\omega
\ , \label{pq1w}
\end{equation}
where a suitable probability vector $\omega$ has to be found.

For any pair of observables, the vectors $\omega$ and
$\omega^{\prime}$ can be obtained within the following procedure.
Let $d$-dimensional probabilistic vectors $p$ and $q$ be described
according to (\ref{pidf}) and (\ref{qjdf}), respectively. It was
shown in \cite{rpz14} that these probabilistic vectors obey
(\ref{pq1w}) with
\begin{equation}
\omega=(s_{1},s_{2}-s_{1},\ldots,s_{d}-s_{d-1})
\ . \label{wdd}
\end{equation}
The sequence of positive elements $s_{k}$ is determined by
(\ref{subvk}) and (\ref{skdf}) with the unitary matrix
$\wm=\bigl[\bigl[\langle{e}_{i}|f_{j}\rangle\bigr]\bigr]$.
Strictly speaking, the integer subscript $k$ in (\ref{skdf})
ranges from $1$ up to $d^{2}-1$. It is easy to see, however, that
$s_{d}=1$ in the case of two orthonormal bases. Next numbers of
the sequence $\{s_{k}\}$ will be $1$ as well. So, we have
discarded several zeros from the right-hand side of (\ref{wdd}).
An earlier result derived in \cite{prz13} can be written as
(\ref{pq0w}), where
\begin{equation}
\omega^{\prime}=(t_{1},t_{2}-t_{1},\ldots,t_{d}-t_{d-1})
\ ,
\qquad
t_{k}=\frac{(1+s_{k})^{2}}{4}
\ . \label{wpdd}
\end{equation}

In principle, majorization relations (\ref{pq0w}) and (\ref{pq1w})
already impose some restrictions on the probabilistic vectors $p$
and $q$. Moreover, they can be easily converted into entropic
uncertainty relations. As the R\'{e}nyi entropy is Schur concave,
the majorization relation (\ref{pq0w}) leads to the following
bounds for R\'{e}nyi entropy with $\alpha>0$ \cite{fgg13,prz13},
\begin{equation}
H_{\alpha}(p)+H_{\alpha}(q)\geq{H}_{\alpha}(\omega^{\prime})
\ . \label{oldmr}
\end{equation}
The authors of \cite{fgg13} also used majorization relations of
the tensor-product type. They formulate an optimization problem
for finding a majorizing vector in (\ref{pq0w}). This problem can
be easily extended to the case of several POVM measurements.
However, since no general, effective algorithm for solving the
optimization problem is available, in this work we extend other
techniques applied for orthogonal measurements in \cite{prz13}.

As was shown in \cite{rpz14}, majorization relation (\ref{pq1w})
allows one to improve entropic bounds. For $0<\alpha\leq1$, we
have
\begin{equation}
H_{\alpha}(p)+H_{\alpha}(q)\geq{H}_{\alpha}(\omega)
\ . \label{nwmr0}
\end{equation}
This bound is stronger, since $\omega\prec\omega^{\prime}$ and,
herewith, $H_{\alpha}(\omega)\geq{H}_{\alpha}(\omega^{\prime})$
\cite{rpz14}. For $\alpha>1$  relation (\ref{nwmr0}) does not
hold. However, in the case $\alpha>1$ the sum of two R\'{e}nyi
entropies satisfies another inequality \cite{rpz14}
\begin{equation}
H_{\alpha}(p)+H_{\alpha}(q)\geq
\frac{2}{1-\alpha}{\>}
\ln\!\left(
\frac{1}{2}+\frac{1}{2}\,\sum\nolimits_{i=1}^{d}\omega_{i}^{\alpha}
\right)
\, . \label{nwmr1}
\end{equation}
It turned out that the sum of two Tsallis $\alpha$-entropies is
bounded from below similarly to (\ref{nwmr0}). For
any $\alpha>0$ we have
\begin{equation}
T_{\alpha}(p)+T_{\alpha}(q)\geq{T}_{\alpha}(\omega)
\ . \label{nwmr01}
\end{equation}
Extensions of the above majorization relations to several
orthonormal bases are discussed in \cite{prz13,rpz14}. We aim to
generalize majorization relations in another direction connected
with quantum operations.

\section{Majorization uncertainty relations for two quantum operations}\label{sec4}

In this section, we formulate majorization uncertainty relations
for a pair of trace-preserving quantum operations. For simplicity,
we focus on the case of the same dimensionality of the input and
output states. Reformulation for different dimensions of the input
and output spaces is a purely technical task. Since block matrices
will be used in our consideration, dimensionality should be
mentioned explicitly. As above, the Hilbert space of interest is
referred to as $\hh_{d}$.

Let $\Phi:{\>}\lnp(\hh_{d})\rightarrow\lnp(\hh_{d})$ be a
trace-preserving completely positive (TPCP) map with Kraus operators
$\am_{i}$ of unraveling $\cla$. Let
$\Psi:{\>}\lnp(\hh_{d})\rightarrow\lnp(\hh_{d})$ be another
trace-preserving completely positive map with Kraus operators
$\bn_{j}$ of unraveling $\clb$. Adding zero operators if
necessary, we can assume that each of the unravelings has $N$
Kraus operators. Let $|\psi\rangle\in\hh_{d}$ denote the input
state, so that probabilities of a given measurement outcome read
in both cases
\begin{equation}
p_{i}(\cla|\psi)=\langle\psi|\am_{i}^{\dagger}\am_{i}|\psi\rangle
\ , \qquad
q_{j}(\clb|\psi)=\langle\psi|\bn_{j}^{\dagger}\bn_{j}|\psi\rangle
\ . \label{prpq}
\end{equation}
For the initial state $\bro\in\lsp(\hh_{d})$, the analogous
equations take the form
\begin{equation}
p_{i}(\cla|\bro)=\Tr\bigl(\am_{i}^{\dagger}\am_{i}\bro\bigr)
\ , \qquad
q_{j}(\clb|\bro)=\Tr\bigl(\bn_{j}^{\dagger}\bn_{j}\bro\bigr)
\ . \label{prpq2}
\end{equation}
The sum of prescribed particular probabilities can be written in a
matrix form. We define column block matrices with $m$, $n$, and
$(m+n)$ blocks of size $d$ as
\begin{equation}
\cn_{Am}:=
\begin{pmatrix}
\am_{1} \\
\cdots \\
\am_{m}
\end{pmatrix}
 , \qquad
\cn_{Bn}:=
\begin{pmatrix}
\bn_{1} \\
\cdots \\
\bn_{n}
\end{pmatrix}
 , \qquad
\gm:=
\begin{pmatrix}
\cn_{Am} \\
\cn_{Bn}
\end{pmatrix}
 . \label{cldf}
\end{equation}
To simplify notation, the collections of indices
$\{i_{1},\ldots,i_{m}\}$ and $\{j_{1},\ldots,j_{n}\}$ are written
as $\{1,\ldots,m\}$ and $\{1,\ldots,n\}$, respectively. The
corresponding block columns are defined by an obvious substitution
and used in further calculations. The columns of all the Kraus
operators are denoted as
\begin{equation}
\cn_{AN}:=
\begin{pmatrix}
\am_{1} \\
\cdots \\
\am_{N}
\end{pmatrix}
 , \qquad
\cn_{BN}:=
\begin{pmatrix}
\bn_{1} \\
\cdots \\
\bn_{N}
\end{pmatrix}
 . \label{cNdf}
\end{equation}
We shall now prove a key result of our approach to uncertainty
relations for an arbitrary pair of quantum operations.

\newtheorem{lem1}[lem01]{Proposition}
\begin{lem1}\label{pn1}
For all $m,n\in\{1,\ldots,N\}$ and arbitrary density matrix
$\bro\in\lsp(\hh_{d})$, we have
\begin{equation}
\sum\nolimits_{i=1}^{m}p_{i}(\cla|\bro)+
\sum\nolimits_{j=1}^{n}q_{j}(\clb|\bro)\leq
1+\|\cn_{Am}\cn_{Bn}^{\dagger}\|_{\infty}
\ , \label{rlm10}
\end{equation}
where the probabilities are defined in (\ref{prpq2}).
Moreover, the following relation holds,
\begin{equation}
\underset{\bro}{\max}
\left(
\sum\nolimits_{i=1}^{m}p_{i}(\cla|\bro)+
\sum\nolimits_{j=1}^{n}q_{j}(\clb|\bro)
\right)
=1+\|\cn_{Am}\cn_{Bn}^{\dagger}\|_{\infty}
\ . \label{rlm11}
\end{equation}
\end{lem1}

{\bf Proof.} The indices $m$ and $n$ are fixed in the calculations
of the proof. Using the spectral decomposition, we have
\begin{equation}
p_{i}(\cla|\bro)=\sum\nolimits_{\lambda}\lambda\>{p}_{i}(\cla|\psi_{\lambda})
\ , \label{sdcc}
\end{equation}
where the vectors $|\psi_{\lambda}\rangle$ form an eigenbasis of
$\bro$. Due to (\ref{sdcc}), we prove (\ref{rlm10}) for
pure input states. One has
\begin{equation}
\sum\nolimits_{i=1}^{m} p_{i}(\cla|\psi)+
\sum\nolimits_{j=1}^{n} q_{j}(\clb|\psi)=
\langle\psi|\gm^{\dagger}\gm|\psi\rangle
\ . \label{smpr}
\end{equation}
Due to the properties of the spectral norm, we obtain
\begin{equation}
\langle\psi|\gm^{\dagger}\gm|\psi\rangle
\leq\|\gm^{\dagger}\gm\|_{\infty}
=\|\gm\gm^{\dagger}\|_{\infty}
\ . \label{prmcc}
\end{equation}
Reexpressing the block matrix $\gm\gm^{\dagger}$, we further
write
\begin{equation}
\gm\gm^{\dagger}=
\begin{pmatrix}
\cn_{Am}\cn_{Am}^{\dagger} & \zmx \\
\zmx & \cn_{Bn}\cn_{Bn}^{\dagger}
\end{pmatrix}
+\begin{pmatrix}
\zmx & \cn_{Am}\cn_{Bn}^{\dagger} \\
\cn_{Bn}\cn_{Am}^{\dagger} & \zmx
\end{pmatrix}
 . \label{gmmg}
\end{equation}
By $\zmx$, we mean zero blocks of the corresponding size. Writing
the formula
\begin{equation}
\begin{pmatrix}
\zmx & \ay \\
\ay^{\dagger} & \zmx
\end{pmatrix}
\begin{pmatrix}
\zmx & \ay \\
\ay^{\dagger} & \zmx
\end{pmatrix}
=\begin{pmatrix}
\ay\ay^{\dagger} & \zmx \\
\zmx & \ay^{\dagger}\ay
\end{pmatrix}
 , \label{ydyb}
\end{equation}
we see that the spectral norm of the Hermitian matrix in the
left-hand side of (\ref{ydyb}) is equal to $\|\ay\|_{\infty}$.
Due to properties of the spectral norm, including the triangle
inequality, we obtain
\begin{equation}
\|\gm\gm^{\dagger}\|_{\infty}\leq
\max\left\{
\|\cn_{Am}\|_{\infty}^{2},\|\cn_{Bn}\|_{\infty}^{2}
\right\}+\|\cn_{Am}\cn_{Bn}^{\dagger}\|_{\infty}
\ . \label{cacb}
\end{equation}
Further, we have $\|\cn_{Am}\|_{\infty}\leq\|\cn_{AN}\|_{\infty}$
and $\|\cn_{Bn}\|_{\infty}\leq\|\cn_{BN}\|_{\infty}$ by
submultiplicativity of the spectral norm. For a trace-preserving
quantum operation, the Kraus operators obey the identity resolution
(\ref{prtr}). That is, we have
\begin{equation}
\cn_{AN}^{\dagger}\cn_{AN}=
\cn_{BN}^{\dagger}\cn_{BN}=\pen_{d}
\ . \label{cacbp}
\end{equation}
Hence, non-zero singular values of $\cn_{AN}$ and $\cn_{BN}$ are
all $1$. Substituting these facts to (\ref{prmcc}) finally gives
\begin{equation}
\sum\nolimits_{i=1}^{m} p_{i}+
\sum\nolimits_{j=1}^{n} q_{j}
\leq
1+\|\cn_{Am}\cn_{Bn}^{\dagger}\|_{\infty}
\leq
1+\|\cn_{AN}\cn_{BN}^{\dagger}\|_{\infty}
\ . \label{cbnd}
\end{equation}
This completes the proof of (\ref{rlm10}). As is seen from
(\ref{prmcc}), inequality (\ref{rlm10}) is saturated with
those eigenvector of $\gm^{\dagger}\gm$ that corresponds to the
maximal eigenvalue. Hence, the claim (\ref{rlm11}) follows.
$\blacksquare$

This bound holds for all pure input states, and therefore, for any
mixed state. We are now ready to formulate majorization
uncertainty relations for two trace-preserving quantum operations.
The structure of (\ref{rlm11}) gives us a hint for an appropriate
extension of the definition (\ref{subvk}). We will deal with
rectangular matrices whose entries are matrices again. The size of
such entries is fixed and determined by the dimensionality. The
main result is posed as follows.

\newtheorem{nepr}[lem01]{Proposition}
\begin{nepr}\label{newp}
Let $\cla=\{\am_{i}\}$ and $\clb=\{\bn_{j}\}$ be unravelings of
two TPCP maps. Let us define $\omega$ and $\omega^{\prime}$
according to (\ref{wdd}) and (\ref{wpdd}), where the numbers
(\ref{skdf}) should be replaced with the numbers (\ref{bskdf})
calculated for the block matrix
\begin{equation}
\ax=\cn_{AN}\cn_{BN}^{\dagger}
\, , \qquad
\ax_{ij}=\am_{i}\bn_{j}^{\dagger}
\, . \label{nereq}
\end{equation}
Then the probabilistic vectors $p(\cla|\bro)$ and $q(\clb|\bro)$
defined due to (\ref{prpq2}) obey the majorization relations
(\ref{pq0w}) and (\ref{pq1w}).
\end{nepr}

The justification of the claim is quite similar to the reasons
given for (\ref{pq1w}) in \cite{rpz14} and for (\ref{pq0w}) in
\cite{prz13}. The key point has been proved above as Proposition
\ref{pn1}. Formally, we merely replace $s_{k}$ with $c_{k}$ in
both the equations (\ref{wdd}) and (\ref{wpdd}). It should be
noted that the sequence $\{c_{k}\}$ contains $N^{2}-1$ numbers.
With certainty, the final number of this sequence is equal to $1$
due to (\ref{cacbp}). For two quantum operations, each with $N$
Kraus operators, the majorizing vectors $\omega$ and
$\omega^{\prime}$ are generally comprised of $N^{2}-1$ elements.
Thus, expressions (\ref{wdd}) and (\ref{wpdd}) should be
recast accordingly.

Possible values of spectral norms dealt with in (\ref{bskdf}) have
a certain impact on majorization relations. Let us consider some
properties of such norms. In general, elements of the sequence
$\{c_{k}\}$ depend on the choice of Kraus operators. Recall that
the Kraus operators are treated as blocks forming  block matrices
(\ref{cldf}). We shall now prove that the freedom in operator-sum
representation does not alter the quantity
$\|\cn_{AN}\cn_{BN}^{\dagger}\|_{\infty}$.

\newtheorem{lem2}[lem01]{Lemma}
\begin{lem2}\label{pn2}
Let the sets $\{\am_{i}\}$ and $\{\widetilde{\am}_{i}\}$ be
unravelings of the TPCP map $\Phi$, and let the set $\{\bn_{i}\}$
be an unraveling of the TPCP map $\Psi$. Then,
\begin{equation}
\|\widetilde{\cn}_{AN}\cn_{BN}^{\dagger}\|_{\infty}=\|\cn_{AN}\cn_{BN}^{\dagger}\|_{\infty}
\ , \label{rlm2}
\end{equation}
where the matrices $\cn_{AN}$ and $\cn_{BN}$ are defined in
(\ref{cNdf}), and
\begin{equation}
\widetilde{\cn}_{AN}:=
\begin{pmatrix}
\widetilde{\am}_{1} \\
\cdots \\
\widetilde{\am}_{N}
\end{pmatrix}
 . \label{wcNa}
\end{equation}
\end{lem2}

{\bf Proof.} Two unravelings of the same map are linked by the
relation (\ref{eqvun}) with a fixed unitary
$\gmsf_{A}\in\mset_{N}(\zset)$. Therefore, we can represent the
matrix (\ref{wcNa}) in the form
\begin{equation}
\widetilde{\cn}_{AN}=(\gmsf_{A}\otimes\pen_{d}){\,}\cn_{AN}
\, . \label{wcNaa}
\end{equation}
Since the spectral norm is unitarily invariant, its value is the
same for matrices $\widetilde{\cn}_{AN}\cn_{BN}^{\dagger}$ and
$\cn_{AN}\cn_{BN}^{\dagger}$. $\blacksquare$

Suppose that the Kraus operators $\widetilde\bn_{i}$ give another
representation of the map $\Psi$. Similarly to (\ref{wcNaa}), we
have
\begin{equation}
\widetilde{\cn}_{BN}=(\gmsf_{B}\otimes\pen_{d}){\,}\cn_{BN}
\, . \label{wcNbb}
\end{equation}
Applying the above result, we see that the spectral norm
$\|\cn_{AN}\widetilde{\cn}_{BN}^{\dagger}\|_{\infty}$ is equal
to the right-hand side of (\ref{rlm2}).

On the other hand, spectral norms of matrices of the form
$\cn_{Am}\cn_{Bn}^{\dagger}$ generally depend on the choice of Kraus
operators. This is a reflection of the fact that the given TPCP
map is characterized by an entire family of unravelings.

\newtheorem{lem3}[lem01]{Lemma}
\begin{lem3}\label{pn3}
Let $\cn_{Am}$ be a column of $m$ blocks $\am_{i}$, and let
$\cn_{Bn}$ be a column of $n$ blocks $\bn_{j}$. Then the following
equality holds
\begin{equation}
\|\cn_{Am}\cn_{Bn}^{\dagger}\|_{\infty}=
\bigl\|\cn_{|A|m}\cn_{|B|n}^{\dagger}\bigr\|_{\infty}
\ . \label{rlm3}
\end{equation}
Here, the column block matrices $\cn_{|A|m}$ and $\cn_{|B|n}$
are defined as
\begin{equation}
\cn_{|A|m}:=
\begin{pmatrix}
|\am_{1}| \\
\cdots \\
|\am_{m}|
\end{pmatrix}
 , \qquad
\cn_{|B|n}:=
\begin{pmatrix}
|\bn_{1}| \\
\cdots \\
|\bn_{n}|
\end{pmatrix}
 , \label{dfcl}
\end{equation}
in terms of positive operators
$|\am_{i}|:=\bigl(\am_{i}^{\dagger}\am_{i}\bigr)^{1/2}$,
$|\bn_{j}|:=\bigl(\bn_{j}^{\dagger}\bn_{j}\bigr)^{1/2}$.
\end{lem3}

{\bf Proof.} Using the polar decomposition, we can write
\begin{equation}
\am_{i}=\um_{i}{\,}|\am_{i}|
\ , \qquad
\bn_{i}=\vm_{j}{\,}|\bn_{j}|
\ , \label{abpd}
\end{equation}
where $\um_{i}$ and $\vm_{j}$ are some unitary matrices. Let us
introduce the diagonal matrices of unitary blocks:
\begin{equation}
\um_{Am}:={\mathrm{diag}}(\um_{1},\ldots,\um_{m})
\ , \qquad
\vm_{Bn}:={\mathrm{diag}}(\vm_{1},\ldots,\vm_{n})
\ . \label{umvn}
\end{equation}
Due to (\ref{abpd}) we easily obtain
\begin{equation}
\cn_{Am}=\um_{Am}\cn_{|A|m}
\ , \qquad
\cn_{Bn}=\vm_{Bn}\cn_{|B|n}
\ , \label{cabmn}
\end{equation}
whence
\begin{equation}
\cn_{Am}\cn_{Bn}^{\dagger}=
\um_{Am}\cn_{|A|m}\cn_{|B|n}^{\dagger}\vm_{Bn}^{\dagger}
\ . \label{abcmn}
\end{equation}
As the spectral norm is unitarily invariant, the last formula
leads to (\ref{rlm3}). $\blacksquare$

The above statements describe some general properties of the
quantities of interest. Thus, there are many particular forms of
an explicit formulation of majorization relations for two quantum
operations. It is connected with the well-known freedom in the
choice of Kraus operators.

\section{Simple examples of majorization relations}\label{sec5}

In this section, we will exemplify the developed scheme in the
simplest non-trivial case. It is useful to consider two qubit
quantum operations, each with two Kraus operators. The general
formulas for elements of the sequence $\{c_{k}\}$ can be reduced
to
\begin{align}
c_{1}&=\underset{ij}{\max}\,\|\am_{i}\bn_{j}^{\dagger}\|_{\infty}
\ , \label{s1cl}\\
c_{2}&=\max\bigl\{\|\am_{+}\|_{\infty},\|\am_{-}\|_{\infty},\|\bn_{+}\|_{\infty},\|\bn_{-}\|_{\infty}\bigr\}
\ , \label{s2cl}\\
c_{3}&=1
\ . \label{s3cl}
\end{align}
The result (\ref{s3cl}) follows from (\ref{cacbp}), since
$N=2$ here. Expression (\ref{s2cl}) is justified as follows.
For convenience, we will further denote labels by $\pm$. The maps
are written as
\begin{align}
\Phi(\bro)&=\sum\nolimits_{i=\pm}\am_{i}\bro\,\am_{i}^{\dagger}
\ , \label{ampm}\\
\Psi(\bro)&=\sum\nolimits_{j=\pm}\bn_{j}\bro\,\bn_{j}^{\dagger}
\ . \label{bnpm}
\end{align}
The following observations can be made from (\ref{ampm}) and
(\ref{bnpm}). First of all, we  deal with the positive operators
$|\am_{\pm}|$ and $|\bn_{\pm}|$. As the number of Kraus operators
is two, the operators $|\am_{+}|$ and $|\am_{-}|$ have the common
eigenbasis. Hence, we can represent the pair $|\am_{\pm}|$ in
terms of the Bloch vector and another parameter connected with the
traces. The pair $|\bn_{\pm}|$ is treated similarly.

Unfortunately, obtained expressions for the norms are more
involved in comparison to (\ref{s1cl})--(\ref{s3cl}). However,
they become considerably simpler, is the Kraus operators are all
normalized to unity in sense of the Hilbert--Schmidt inner
product. Therefore we assume that
\begin{equation}
\Tr\bigl(\am_{\pm}^{\dagger}\am_{\pm}\bigr)=
\Tr\bigl(\bn_{\pm}^{\dagger}\bn_{\pm}\bigr)=1
\  \label{trab1}
\end{equation}
and focus on this example, since it covers the
previously known case of two orthogonal measurements.
Keeping the completeness relation, we can now write
\begin{align}
\am_{\pm}^{\dagger}\am_{\pm}=|\am_{\pm}|^{2}=\frac{1}{2}\,
\bigl(
\pen_{2}\pm\vec{a}\cdot\vec{\bau}
\bigr)
\, , \label{ama}\\
\bn_{\pm}^{\dagger}\bn_{\pm}=|\bn_{\pm}|^{2}=\frac{1}{2}\,
\bigl(
\pen_{2}\pm\vec{b}\cdot\vec{\bau}
\bigr)
\, , \label{bnb}
\end{align}
where $\vec{\bau}=(\bau_{x},\bau_{y},\bau_{z})$ denote the vector of
three Pauli matrices. The length of any of two Bloch vectors
$\vec{a}$ and $\vec{b}$ is not larger than $1$. Denoting
$a=|\vec{a}|$ and $b=|\vec{b}|$, we easily obtain
\begin{equation}
c_{2}=\sqrt{\frac{1+\max\{a,b\}}{2}}
\ . \label{s2ab}
\end{equation}
Slightly more complicated calculation of $c_{1}$ gives
 the  following expression,
\begin{equation}
c_{1}^{2}=\frac{1}{4}
\left(
1+|\vec{a}\cdot\vec{b}|
+\sqrt{\bigl(1+|\vec{a}\cdot\vec{b}|\bigr)^{2}-(1-a^{2})(1-b^{2})}
\right)
\, . \label{s12ab}
\end{equation}
As mentioned in (\ref{s3cl}), we have $c_{3}=1$. We can
now write the majorization relations (\ref{pq1w}) and (\ref{pq0w})
for two quantum operations with
\begin{align}
\omega&=(c_{1},c_{2}-c_{1},1-c_{2})
\ , \label{omb} \\
\omega^{\prime}&=(t_{1},t_{2}-t_{1},1-t_{2})
\ , \label{omt}
\end{align}
where $t_{k}=(1+c_{k})^{2}/4$. Entropic uncertainty relations are
then expressed as (\ref{oldmr})--(\ref{nwmr01}) with the
corresponding conditions on the entropic parameter $\alpha$.

\begin{figure}
\includegraphics[width=9.0cm]{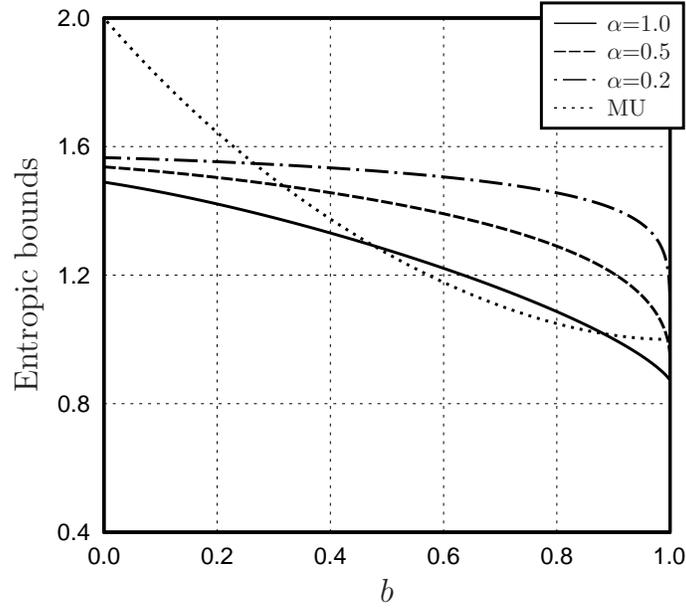}
\caption{\label{fig1} Entropic uncertainty relations for two
quantum operations (\ref{bnb}) shown as a function of the length
$b$ of the Bloch vector determining one map for
$\measuredangle(\vec{a},\vec{b})=\pi/2$ and rescaled to the
logarithm with base $2$. Majorization bound (\ref{nwmr0}) for
Shannon entropy, ($\alpha=1$, solid line) can be compared with the
Maassen--Uffink like bound (\ref{renuim1}) represented by a dotted
line. For reference, majorization bounds for R{\'e}nyi entropy with
$\alpha=0.5$ (dashed line) and $\alpha=0.2$ (dash-dotted line) are
presented.}
\end{figure}

The described example is helpful for comparing our majorization
bounds with the Maassen--Uffink uncertainty relation. The
Maassen--Uffink result \cite{maass} is perhaps the best known
entropic formulation of the uncertainty principle. It was inspired
by a previous conjecture of Kraus \cite{kraus87}. Although the
Maassen--Uffink uncertainty relation was originally written for
two non-degenerate observables, it can be extended in several
directions \cite{hall97,birula3,massar07,rast102,rast104}. Its
analogue for two generalized measurements is due to Krishna and
Parthasarathy \cite{KP902}. Here we will adopt a later formulation
presented in \cite{rast104}. Let $\ppc=\{\ppm_{i}\}$ and
$\qpc=\{\qpm_{j}\}$ be two positive operator-valued measures
(POVMs) on $\hh_{d}$. In other words, each of these sets contains
positive operators such that
\begin{equation}
\sum\nolimits_{i} \ppm_{i}=
\sum\nolimits_{j} \qpm_{j}=\pen_{d}
\ . \label{cpenr}
\end{equation}
If the pre-measurement state is described by density matrix
$\bro$, we correspondingly deal with probabilities
\begin{equation}
p_{i}(\ppc|\bro)=\Tr(\ppm_{i}\bro)
\ , \qquad
q_{j}(\qpc|\bro)=\Tr(\qpm_{j}\bro)
\ . \label{ppijq}
\end{equation}
We also introduce the function
\begin{equation}
\bar{f}(\ppc,\qpc):=
\max\Bigl\{\bigl\|\ppm_{i}^{1/2}\qpm_{j}^{1/2}\bigr\|_{\infty}:
\> \ppm_i\in\ppc,\> \qpm_j\in\qpc\Bigr\}
\, .
\end{equation}
Calculating with these probabilities leads to the entropies
$H_{\alpha}(\ppc|\bro)$ and $H_{\beta}(\qpc|\bro)$ according to
the definition (\ref{renent}). For $1/\alpha+1/\beta=2$, the
R\'{e}nyi entropies satisfy the state-independent bound
\begin{equation}
H_{\alpha}(\ppc|\bro)+H_{\beta}(\qpc|\bro)\geq-2\ln\bar{f}(\ppc,\qpc)
\ . \label{renuim}
\end{equation}
In particular, the sum of
Shannon entropies is bounded from below as
\begin{equation}
H_{1}(\ppc|\bro)+H_{1}(\qpc|\bro)\geq-2\ln\bar{f}(\ppc,\qpc)
\ . \label{renuim1}
\end{equation}
The condition $1/\alpha+1/\beta=2$ reflects the fact that the
Maassen--Uffink result is based on Riesz's theorem (see, e.g.,
theorem 297 in the book \cite{hardy}). An alternative viewpoint is
that uncertainty relations follow from the monotonicity of the
quantum relative entropy \cite{ccyz12}. Then the condition on
$\alpha$ and $\beta$ is due to the so-called duality of entropies
\cite{ccyz12}.

\begin{figure}
\includegraphics[width=9.0cm]{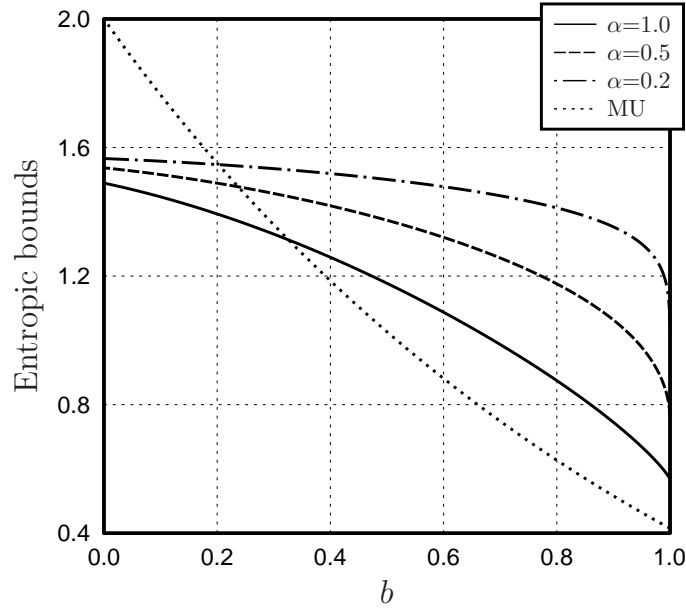}
\caption{\label{fig2}
Entropic uncertainty relations as in Fig. \ref{fig1}
for $\measuredangle(\vec{a},\vec{b})=\pi/3$.
Note a larger range of parameter $b$
for which majorization bound (\ref{nwmr0})
is stronger than (\ref{renuim1}).
}
\end{figure}

Majorization relations (\ref{pq1w}) and (\ref{pq0w}) lead to
uncertainty bounds on the sum of corresponding entropies of the
same order. Except for (\ref{renuim1}), the Maassen--Uffink
approach provides a lower bound on the sum of entropies of
different orders. Relaxing a restriction, the right-hand side of
(\ref{renuim1}) gives a lower bound on the sum of two R\'{e}nyi's
entropies of the same order $0<\alpha\leq1$. It is known that the
R\'{e}nyi $\alpha$-entropy does not increase with growth of the
parameter $\alpha$.

We shall now return to our example of two quantum operations with
the Kraus operators (\ref{ama}) and (\ref{bnb}), respectively. Let
us denote
\begin{equation}
\ppm_{\pm}=|\am_{\pm}|^{2}
\ , \qquad
\qpm_{\pm}=|\bn_{\pm}|^{2}
\ . \label{pqab}
\end{equation}
One then gives $\bar{f}(\ppc,\qpc)=c_{1}$ in the sense of
(\ref{s1cl}). Here, we can  replace $\am_{i}$ and
$\bn_{j}^{\dagger}$ with $|\am_{i}|$ and $|\bn_{j}|$ as the
spectral norm is unitarily invariant. Thus, for $0<\alpha\leq1$ we
write
\begin{equation}
H_{\alpha}(p)+H_{\alpha}(q)\geq-2\ln{c}_{1}
\ , \label{mufn}
\end{equation}
where $p_{\pm}=\Tr(\ppm_{\pm}\bro)$ and
$q_{\pm}=\Tr(\qpm_{\pm}\bro)$. The inequality (\ref{mufn}) is
herewith R\'{e}nyi's formulation of the Maassen--Uffink type for
the example of two qubit channels. Another approach to derive
R\'{e}nyi-entropy uncertainty relations was developed in
\cite{zbp2014}. We aim to compare (\ref{mufn}) with (\ref{nwmr0}),
which follows from relation (\ref{pq1w}) of the direct-sum type.

Let us take two vectors $\vec{a}$ and $\vec{b}$ of the same length
$b=|\vec{a}|=|\vec{b}|$. We then write $c_{2}=\sqrt{(1+b)/2}$ and
$|\vec{a}\cdot\vec{b}|=b^{2}\cos\measuredangle(\vec{a},\vec{b})$.
The result for $c_{1}$ is obtained by substituting the scalar
product into the right-hand side of (\ref{s12ab}). For the given
$0<\alpha\leq1$, the majorization-based entropic bound then
follows from (\ref{nwmr0}) by substituting (\ref{omb}). Running
$b$, we will visualize the bounds of interest. For two outcomes,
the maximal value of (\ref{renent}) is equal to $\ln2$. It is
convenient to rescale R\'{e}nyi's entropies by $\ln2$ so the
logarithm becomes taken in the base $2$.

The right-hand side of (\ref{nwmr0}) is shown in Fig. \ref{fig1}
for $\measuredangle(\vec{a},\vec{b})=\pi/2$ and for
three values of the parameter $\alpha=1$, $\alpha=0.5$, and $\alpha=0.2$.
There is a range of $b$, in which formula (\ref{nwmr0}) gives a
stronger bound. At the same time, the Maassen--Uffink bound is
better for small $b$, as the probability distributions become
almost uniform. In Fig. \ref{fig2}, the right-hand side of
(\ref{nwmr0}) is shown for
$\measuredangle(\vec{a},\vec{b})=\pi/3$. In this case the range of
parameter, for which  the bound (\ref{nwmr0}) is stronger becomes
greater and the difference between both bounds becomes larger. On
the other hand, the Maassen--Uffink bound is still better for
sufficiently small $b$. With further decrease of the angle
$\measuredangle(\vec{a},\vec{b})$, the picture is similar to this
shown in Fig. \ref{fig2}. For sufficiently large $b$, relation
(\ref{nwmr0}) provides a significant improvement with respect to
the the Maassen--Uffink relation.

\section{Equivalence in the case of two orthonormal bases}\label{sec6}

In this section, we show that the presented formulation of
majorization uncertainty relations includes earlier results of
\cite{prz13,rpz14} as particular cases. Let us begin by
simplifying the above example of qubit channels with the Kraus
operators (\ref{ama}) and (\ref{bnb}). The case of two orthonormal
bases is realized with two projective measurements of a qubit. To
avoid a confusing notation, the corresponding unit vectors will be
denoted by $\vec{m}$ and $\vec{n}$. Instead of the Kraus operators
(\ref{ama}) and (\ref{bnb}), we now deal with rank-one projectors
\begin{align}
\psf_{\pm}&=|e_{\pm}\rangle\langle{e}_{\pm}|=
\frac{1}{2}\,
\bigl(
\pen_{2}\pm\vec{m}\cdot\vec{\bau}
\bigr)
\, , \label{psdf}\\
\lsf_{\pm}&=|f_{\pm}\rangle\langle{f}_{\pm}|=
\frac{1}{2}\,
\bigl(
\pen_{2}\pm\vec{n}\cdot\vec{\bau}
\bigr)
\, . \label{lsdf}
\end{align}
Here, the vectors of each bases are numbered in the ``plus/minus''
notation. For $i,j=\pm1$, simple calculations give
\begin{equation}
\Tr(\psf_{i}\lsf_{j})=
\bigl|\langle{e}_{i}|f_{j}\rangle\bigr|^{2}
=\frac{1+ij{\,}\vec{m}\cdot\vec{n}}{2}
\ . \label{ijmn}
\end{equation}
In our example, the formulas (\ref{s12ab}) and (\ref{ijmn}) lead
to the same result, since
\begin{equation}
s_{1}=\underset{ij}{\max}\bigl|\langle{e}_{i}|f_{j}\rangle\bigr|
=\sqrt{\frac{1+|\vec{m}\cdot\vec{n}|}{2}}
=c_{1}
\ . \label{s1mn}
\end{equation}
In the case considered, the formula (\ref{s2ab}) further gives
$c_{2}=1$. From (\ref{skdf}), we obtain the same value.
Indeed, the quantity $s_{2}$ is the maximum among the spectral
norms of the matrices
\begin{equation}
\begin{pmatrix}
\langle{e}_{i}|f_{+}\rangle & \langle{e}_{i}|f_{-}\rangle
\end{pmatrix}
\, , \qquad
\begin{pmatrix}
\langle{e}_{+}|f_{j}\rangle \\
\langle{e}_{-}|f_{j}\rangle
\end{pmatrix}
\, . \label{efij}
\end{equation}
In two dimensions, the spectral norm of each of the matrices
(\ref{efij}) is equal to $1$. It follows from the completeness
relation. As noted in (\ref{s3cl}), we have $c_{3}=1$. The
spectral norm of a unitary matrix is also $1$, so that $s_{3}=1$.
For two rank-one projective measurements of a qubit, more general approach of
Section \ref{sec4} leads to the same majorization relations as the
previous formulation proposed in \cite{prz13,rpz14}.

We shall now prove the above claim for an arbitrary finite
dimensionality. In such general case, we will assume that the basis
vectors are numbered by the index running from $1$ to $d$. The
precise statement is formulated as follows.

\newtheorem{lem4}[lem01]{Proposition}
\begin{lem4}\label{pn4}
Let $\{|e_{i}\rangle\}$ and $\{|f_{j}\rangle\}$ be two orthonormal
bases in $d$-dimensional Hilbert space. In terms of the rank-one
projectors
\begin{equation}
\psf_{i}=|e_{i}\rangle\langle{e}_{i}|
\ , \qquad
\lsf_{j}=|f_{j}\rangle\langle{f}_{j}|
\ , \label{plef}
\end{equation}
we define the sequence of positive numbers
\begin{equation}
c_{k}:=\max\!\left\{
\|\az\|_{\infty}:{\>}\az\in
\mathcal{BSUB}\bigl(\cn_{\psf{d}}\cn_{\lsf{d}}^{\dagger}{\,},k\bigr)
\right\}
\, . \label{bsplf}
\end{equation}
For all $k=1,\ldots,d$, we then have $c_{k}=s_{k}$, where $s_{k}$
is defined by (\ref{skdf}) with the unitary matrix
$\wm=\bigl[\bigl[\langle{e}_{i}|f_{j}\rangle\bigr]\bigr]$.
\end{lem4}

{\bf Proof.} Let $\um_{e}$ and $\um_{f}$ be the two unitary
transformations such that
\begin{equation}
\um_{e}{\,}|e_{i}\rangle=|i\rangle
\ , \qquad
\um_{f}{\,}|f_{j}\rangle=|j\rangle
\ . \label{umef}
\end{equation}
By $|i\rangle\in\mset_{d\times1}(\zset)$, we mean the column with
$1$ on $i$-th place and with $0$ on all the other places. That is,
the above transformations respectively turn our bases into the
computational one. For all $m\in\{1,\ldots,d\}$, we now write
\begin{equation}
\bigl(\pen_{m}\otimes\um_{e}\bigr){\,}\cn_{\psf{m}}=
{\mathrm{diag}}\underset{\text{$m$ blocks}}{\underbrace{(\um_{e},\ldots,\um_{e})}}
\begin{pmatrix}
\psf_{1} \\
\ldots \\
\psf_{m}
\end{pmatrix}
=
\begin{pmatrix}
|1\rangle\langle{e}_{1}| \\
\ldots \\
|m\rangle\langle{e}_{m}|
\end{pmatrix}
 . \label{muedg}
\end{equation}
Similarly, the block column
$\bigl(\pen_{n}\otimes\um_{f}\bigr){\,}\cn_{\lsf{n}}$ is represented.
As the spectral norm is unitarily invariant, for all
$m,n\in\{1,\ldots,d\}$ we have
\begin{equation}
\bigl\|
\cn_{\psf{m}}\cn_{\lsf{n}}^{\dagger}
\bigr\|_{\infty}=
\left\|
\bigl(\pen_{m}\otimes\um_{e}\bigr)
{\,}\cn_{\psf{m}}\cn_{\lsf{n}}^{\dagger}{\,}
\bigl(\pen_{n}\otimes\um_{f}^{\dagger}\bigr)
\right\|_{\infty}
\, . \label{cplcef}
\end{equation}
Therefore, the numbers of the form (\ref{bsplf}) can be calculated
by replacing $\cn_{\psf{d}}\cn_{\lsf{d}\,}^{\dagger}$ with the
block matrix
\begin{equation}
\widetilde{\wm}:=\bigl(\pen_{d}\otimes\um_{e}\bigr)
{\,}\cn_{\psf{d}}\cn_{\lsf{d}}^{\dagger}{\,}
\bigl(\pen_{d}\otimes\um_{f}^{\dagger}\bigr)
\, . \label{bmdd}
\end{equation}
It follows from (\ref{muedg}) that the $(i,j)$-block of
$\widetilde{\wm}$ is the $d\times{d}$ matrix
$w_{ij}{\,}|i\rangle\langle{j}|$. Due to the structure of the
computational basis, the $(i,j)$-entry of this block reads
$w_{ij}=\langle{e}_{i}|f_{j}\rangle$, while all other entries
are equal zero.

To each block $m\times{n}$ submatrix of $\widetilde{\wm}$, we
assign the $m\times{n}$ submatrix of $\wm$. The latter is obtained
from the former by replacing each $d\times{d}$ block
$w_{ij}{\,}|i\rangle\langle{j}|$ by the single entry $w_{ij}$.
Such a replacement implies that we eliminate some purely zero rows
and columns from the submatrix of blocks. Due to Lemma \ref{pn01},
this operation does not alter the spectral norm. Thus, there is
one-to-one correspondence between block submatrices of
$\widetilde{\wm}$ and usual submatrices of $\wm$ with the
following property. It provides the same value of the spectral
norms of an element of
$\mathcal{BSUB}\bigl(\widetilde{\wm},k\bigr)$ and its ``twin'' of
$\mathcal{SUB}(\wm,k)$. For any given $k=1,\ldots,d$ we actually
deal with the same collection of values of the spectral norm of an
element. Hence, the optimization of (\ref{bsplf}) results in
$c_{k}=s_{k}$, where the latter quantity is defined by
(\ref{skdf}). $\blacksquare$

The statement of Proposition \ref{pn4} justifies that the approach
of Sect. \ref{sec4} is a natural and reasonable extension of the
formulations of the papers \cite{prz13,rpz14}. In the case of two
orthonormal bases, the sequence of numbers $c_{k}$ is actually
reduced to the sequence of numbers $s_{k}$ defined by
(\ref{skdf}). This fact was also exemplified in Sec.~\ref{sec5}
with the one-qubit example.

\section{Concluding Remarks}\label{sec7}

We have derived majorization uncertainty relations for a pair of
trace--preserving quantum operations. They are an extension of the
previous technique developed in \cite{prz13,rpz14}. In the case of
orthonormal bases, majorization relations were posed in terms of
spectral norms of submatrices of a unitary matrix. We have
generalized this approach by considering norms of block matrices
comprised of the corresponding Kraus operators. To any collection
of Kraus operators, we assign a vector of probabilities generated
by the input state. Majorization relations of both the
direct-sum and the tensor-product types are obtained within the
developed scheme. The direct-sum majorization relations for
orthogonal measurements provide bounds stronger than these
obtained with the tensor product of probability vectors
\cite{rpz14}. This conclusion remains valid also in the considered
case of quantum operations.

Existing freedom in the choice of Kraus operators is somehow
reflected in the context of uncertainty relations. First of all,
this choice has an impact on the probabilities assigned to quantum
operations in (\ref{prpq}). The right-hand sides of (\ref{pq1w})
and (\ref{pq0w}) depend also on the chosen Kraus operators. In
this sense, we rather deal with majorization uncertainty relations
for given unravelings of quantum operations, each determined by a
concrete set of measurement operators.

In this work we analyzed the simplest case of two quantum
operations. However, the same approach allows us to get some
non-trivial bounds also in the case of a single operation -- see
\ref{apn1}. As was shown in \cite{rpz14}, majorization
relations of the direct-sum type are naturally reformulated for
several measurements in orthonormal bases. Extending such a
treatment for the case of several quantum operations remains as a
topic of future research.

Entropic majorization relations for two arbitrary quantum
operations derived in this work can be considered as a direct
generalization of previous results concerning orthogonal
measurements \cite{prz13,rpz14}. Dealing with spectral norms of
submatrices, majorization relations of both types
can be obtained in a unified way.
Presented approach is focused on norms of block
submatrices of block matrices comprised of the corresponding Kraus
operators.

Although majorization relations derived in this work were
illustrated with an explicit example for a pair of trace-preserving completely
positive maps acting on a single qubit, they are
valid for arbitrary two quantum operations acting on a
finite-dimensional state. Therefore these results can be applied
for studying various problems in the theory of quantum
information.

As entropic uncertainty relations represent lower bounds for the
average entropy characterizing a given measurement, one studies
also so-called {\sl entropic certainty relations}
\cite{sanchez93}, i.e. the corresponding upper bounds. Although for
two orthogonal measurements in $d$-dimensional space the trivial
upper bound equal to $\log{d}$ is always saturated \cite{KLJR14},
some nontrivial certainty relations for three or more orthogonal
measurements were recently derived in \cite{PRCPZ15}. It would be
therefore interesting to study an analogous problem for quantum
operations. To be more specific, one can pose the problem of
finding  upper bounds for the average entropy characterizing
generalized measurements with the number of Kraus operators equal
to the rank of the Choi matrices corresponding to the investigated
quantum operations.

\acknowledgments

We are grateful to anonymous referees for valuable comments. It is
a pleasure to thank to Zbigniew Pucha{\l}a and {\L}ukasz Rudnicki
for joint work on entropic uncertainty relations and numerous
fruitful discussions. This work was supported by the grant number
DEC-2015/18/A/ST2/00274 financed by the Polish National Science
Centre.

\appendix

\section{Majorization relations for a single quantum operation}\label{apn1}

In this section, we present applications of the majorization
approach to a single quantum operation. Let the set
$\cla=\{\am_{i}\}$ with $N$ elements be unraveling of a TPCP map
$\Phi$. The following statement takes place. For all
$m\in\{1,\ldots,N\}$ and arbitrary density matrix
$\bro\in\lsp(\hh_{d})$, we have
\begin{equation}
\sum\nolimits_{i=1}^{m}p_{i}(\cla|\bro)
\leq\|\cn_{Am}^{\dagger}\cn_{Am}\|_{\infty}
=\|\cn_{Am}\cn_{Am}^{\dagger}\|_{\infty}
\ . \label{arlm10}
\end{equation}
Here, the block matrix $\cn_{Am}$ is defined according to
(\ref{cldf}). Similarly to (\ref{rlm10}), the result
(\ref{arlm10}) follows from the relation
\begin{equation}
\sum\nolimits_{i=1}^{m}
\langle\psi|\am_{i}^{\dagger}\am_{i}|\psi\rangle
\leq\|\cn_{Am}^{\dagger}\cn_{Am}\|_{\infty}
\ , \label{arlm101}
\end{equation}
where $|\psi\rangle$ is an arbitrary pure state. Therefore, the
vector comprised of probabilities $p_{i}(\cla|\bro)$ is majorized
by the vector
\begin{equation}
\wto=\bigl(\wtc_{1},\wtc_{2}-\wtc_{1},\ldots,\wtc_{N^{2}-1}-\wtc_{N^{2}-2}\bigr)
\ , \label{wtodf}
\end{equation}
with entries given by the norms
\begin{equation}
\wtc_{k}:=\max\!\left\{
\|\az\|_{\infty}:{\>}\az\in\mathcal{BSUB}\bigl(\cn_{AN}\cn_{AN}^{\dagger},k\bigr)
\right\}
 , \label{abskdf}
\end{equation}
were notation introduced in (\ref{bsubvk}) is used. Note that for
any $k>1$ the norms are non-decreasing, $c_k\geq c_{k-1}$, so that
the entries of the vector $\wto$ are non-negative. The
completeness relation can be reexpressed as
$\cn_{AN}^{\dagger}\cn_{AN}=\pen_{d}\,$, whence we always have
\begin{equation}
\wtc_{N^{2}-1}=\|\cn_{AN}\cn_{AN}^{\dagger}\|_{\infty}
=\|\cn_{AN}^{\dagger}\cn_{AN}\|_{\infty}=1
\ . \label{wtc1}
\end{equation}
Since the R\'{e}nyi and Tsallis entropies are both Schur concave,
the following inequalities follow  for all $\alpha>0$,
\begin{eqnarray}
H_{\alpha}(p)&\geq{H}_{\alpha}\bigl(\wto\bigr)
\ , \label{hapw}\\
T_{\alpha}(p)&\geq{T}_{\alpha}\bigl(\wto\bigr)
\ . \label{tapw}
\end{eqnarray}
Let us illustrate these entropic relations with the following
example. Consider four Kraus operators acting on a single-qubit
system,
\begin{align}
&\am_{1}=
\begin{pmatrix}
0 & \sqrt{a} \\
0 & 0
\end{pmatrix}
,
&\am_{2}=
\begin{pmatrix}
0 & 0 \\
\sqrt{b} & 0
\end{pmatrix}
, \label{aa12}\\
&\am_{3}={\mathrm{diag}}\bigl(0,\sqrt{1-a}\bigr)
\ ,
&\am_{4}={\mathrm{diag}}\bigl(\sqrt{1-b},0\bigr)
\ , \label{aa34}
\end{align}
where real numbers $a$ and $b$ obey $0\leq{a},b\leq1$. Note that
these Kraus operators form a canonical decomposition
\cite{bengtsson} of the quantum operation, as they are mutually
orthogonal, $\Tr(\am_{i}\am_{j})=0$ for $i\neq{j}$. For $a=b=1/2$
the map reduces to the completely depolarizing channel. Formula
(\ref{wtc1}) gives $\wtc_{15}=1$. However, the above example at
once leads to $\wtc_{2}=1$. The latter can be seen from relations
\begin{equation}
\am_{1}^{\dagger}\am_{1}+\am_{3}^{\dagger}\am_{3}=
{\mathrm{diag}}(0,1)
\ , \qquad
\am_{2}^{\dagger}\am_{2}+\am_{4}^{\dagger}\am_{4}=
{\mathrm{diag}}(1,0)
\ . \label{nepen}
\end{equation}
It is sufficient to use the vector
$\wto=\bigl(\wtc_{1},1-\wtc_{1}\bigr)$, whose first element
\begin{equation}
\wtc_{1}=\max\bigl\{a,1-a,b,1-b\bigr\}
\>. \label{wto1ab}
\end{equation}
Here, we generally deal with the four particular probabilities.
For this unraveling, R\'{e}nyi's $\alpha$-entropy is bounded from
below by the binary entropy
$H_{\alpha}\bigl(\wtc_{1},1-\wtc_{1}\bigr)$, and Tsallis'
$\alpha$-entropy is bounded from below by the binary entropy
$T_{\alpha}\bigl(\wtc_{1},1-\wtc_{1}\bigr)$. In general, such
bounds seem not to be very tight. If $a$ and $b$ are both close
to $1/2$, the binary entropy is close to the maximal value.
Let the input state be such that only two
of the four probabilities are significant. Then the above
majorization relations provide rather precise bounds. It
would be interesting to compare majorization relations with
trade-off relations for a single quantum operation. Trade-off
relations for a single quantum operation were proposed in
\cite{rprz12} and later extended in \cite{rast13a}.


\begin{thebibliography}{80}

\bibitem{wh27}
Heisenberg W 1927 {\it Zeitschrift f\"{u}r Physik} {\bf 43} 172

\bibitem{lahti}
Busch P, Heinonen T, and Lahti P J 2007 {\it Phys. Rep.} {\bf 452} 155

\bibitem{hall99}
Hall M J W 1999 {\it Phys. Rev.} A {\bf 59} 2602

\bibitem{robert}
Robertson H P 1929 {\it Phys. Rev.} {\bf 34} 163

\bibitem{maass}
Maassen H and Uffink J B M 1988 {\it Phys. Rev. Lett.} {\bf 60} 1103

\bibitem{zimba00}
Zimba J 2000 {\it Found. Phys.} {\bf 30} 179

\bibitem{ww10}
Wehner S and Winter A 2010 {\it New. J. Phys.} {\bf 12}, 025009

\bibitem{brud11}
I.~Bia{\l}ynicki-Birula and \L. Rudnicki 2011 Entropic uncertainty relations in quantum physics
{\it Statistical Complexity} (Berlin: Springer) pp 1--34

\bibitem{cbtw15}
Coles P J, Berta M, Tomamichel M, and Wehner S 2015 Entropic uncertainty relations and their applications arXiv:1511.04857 [quant-ph]

\bibitem{hirs}
Hirschman I I 1957 {\it Amer. J. Math.} {\bf 79} 152

\bibitem{beck}
Beckner W 1975 {\it Ann. Math.} {\bf 102} 159

\bibitem{birula1}
Bia{\l}ynicki-Birula I and Mycielski J 1975 {\it Commun. Math. Phys.} {\bf 44} 129

\bibitem{deutsch}
Deutsch D 1983 {\it Phys. Rev. Lett.} {\bf 50} 631

\bibitem{kraus87}
Kraus K 1987 {\it Phys. Rev.} D {\bf 35} 3070

\bibitem{ivan92}
Ivanovic I D 1995 {\it J. Phys. A: Math. Gen.} {\bf 25} L363

\bibitem{sanchez93}
S\'{a}nchez J 1993 {\it Phys. Lett.} A {\bf 173} 233

\bibitem{molm09a}
Wu S, Yu S, and M{\o}lmer K 2009 {\it Phys. Rev.} A {\bf 79} 022320

\bibitem{rastmub}
Rastegin A E 2013 {\it Eur. Phys. J.} D {\bf 67} 269

\bibitem{BCCRR10}
Berta M, Christandl M, Colbeck R, Renes J M, and Renner R 2010 {\it Nature Phys.} {\bf 6} 659

\bibitem{coles14}
Coles P J and Piani M 2014 {\it Phys. Rev.} A {\bf 89} 022112

\bibitem{oppwn10}
Oppenheim J and Wehner S 2010 {\it Science} {\bf 330} 1072

\bibitem{renf13}
Ren L-H and Fan H 2014 {\it Phys. Rev.} A {\bf 90} 052110

\bibitem{rastqip15}
Rastegin A E 2015 {\it Quantum Inf. Process.} {\bf 14} 783

\bibitem{Rud15}
Rudnicki {\L} 2015 {\it Phys. Rev.} A {\bf 91} 032123

\bibitem{prtv11}
Partovi M H 2011 {\it Phys. Rev.} A {\bf 84} 052117

\bibitem{prz13}
Pucha{\l}a Z, Rudnicki {\L}, and \.{Z}yczkowski K 2013 {\it J. Phys. A: Math. Theor.} {\bf 46} 272002

\bibitem{fgg13}
Friedland S, Gheorghiu V, and Gour G 2013 {\it Phys. Rev. Lett.} {\bf 111} 230401

\bibitem{rpz14}
Rudnicki {\L}, Pucha{\l}a Z, and \.{Z}yczkowski K 2014 {\it Phys. Rev.} A {\bf 89} 052115

\bibitem{KP902}
Krishna M and Parthasarathy K R 2002 {\it Sankhy\={a}, Ser.} A {\bf 64} 842

\bibitem{rast104}
Rastegin A E 2011 {\it J. Phys. A: Math. Theor.} {\bf 44} 095303

\bibitem{hiai2014}
Hiai F and Petz D 2014 {\it Introduction to Matrix Analysis and Applications} (Heidelberg: Springer)

\bibitem{hj1990}
Horn R A and Johnson C R 1990 {\it Matrix Analysis} (Cambridge: Cambridge University Press)

\bibitem{watrous1}
Watrous J 2015 {\it Theory of Quantum Information} a draft of book (Waterloo: University of Waterloo)
http://www.cs.uwaterloo.ca/{\textasciitilde}watrous/TQI/

\bibitem{bhatia07}
Bhatia R 2007 {\it Positive Definite Matrices} (Princeton: Princeton University Press)

\bibitem{nielsen}
Nielsen M A and Chuang I L 2000 {\it Quantum Computation and Quantum Information} (Cambridge: Cambridge University Press)

\bibitem{bengtsson}
Bengtsson I and \.{Z}yczkowski K 2006 {\it Geometry of Quantum States: An
Introduction to Quantum Entanglement} (Cambridge: Cambridge University Press)

\bibitem{carm}
Carmichael H J 1993 {\it An Open Systems Approach to Quantum Optics} ({\it Lecture Notes
in Physics} vol 18) (Berlin: Springer)

\bibitem{renyi61}
R\'{e}nyi A 1961 On measures of entropy and information {\it Proc. 4th Berkeley Symposium on Mathematical
Statistics and Probability} (Berkeley, CA: University of California Press) pp 547--61

\bibitem{mdsft13}
M\"{u}ller-Lennert M, Dupuis F, Szehr O, Fehr S, and Tomamichel M 2013 {\it J. Math. Phys.} {\bf 54} 122203

\bibitem{berta15}
Berta M, Seshadreesan K P, and Wilde M M 2015 {\it Phys. Rev.} A {\bf 91} 022333

\bibitem{tsallis}
Tsallis C 1988 {\it J. Stat. Phys.} {\bf 52} 479

\bibitem{sudha14}
Rajagopal A K, Sudha, Nayak A S, and Usha Devi A R 2014 {\it Phys. Rev.} A {\bf 89} 012331

\bibitem{rast15ap}
Rastegin A E 2015 {\it Ann. Phys.} {\bf 355} 241

\bibitem{rast16a}
Rastegin A E 2016 {\it Phys. Rev.} A {\bf 93} 032136

\bibitem{hall97}
Hall M J W 1997 {\it Phys. Rev.} A {\bf 55} 100

\bibitem{birula3}
Bia{\l}ynicki-Birula I 2006 {\it Phys. Rev.} A {\bf 74} 052101

\bibitem{massar07}
Massar S 2007 {\it Phys. Rev.} A {\bf 76} 042114

\bibitem{rast102}
Rastegin A E 2010 {\it J. Phys. A: Math. Theor.} {\bf 43} 155302

\bibitem{hardy}
Hardy G H, Littlewood J E, and Polya G 1934 {\it Inequalities} (London: Cambridge University Press)

\bibitem{ccyz12}
Coles P J, Colbeck R, Yu L, and Zwolak M 2012 {\it Phys. Rev. Lett.} {\bf 108} 210405

\bibitem{zbp2014}
Zozor S, Bosyk G M, and Portesi M 2014 {\it J. Phys. A: Math. Theor.} {\bf 47} 495302

\bibitem{KLJR14}
Korzekwa K, Lostaglio M, Jennings D, and Rudolph T 2014 {\it Phys Rev.} A {\bf 89} 042122

\bibitem{PRCPZ15}
Pucha{\l}a Z, Rudnicki {\L}, Chabuda K, Paraniak M, and \.{Z}yczkowski K 2015 {\it Phys Rev.} A {\bf 92} 032109

\bibitem{rprz12}
Roga W, Pucha{\l}a Z, Rudnicki {\L}, and \.{Z}yczkowski K 2013 {\it Phys Rev.} A {\bf 87} 032308

\bibitem{rast13a}
Rastegin A E 2013 {\it J. Phys. A: Math. Theor.} {\bf 46} 285301


\end{thebibliography}
\end{document}